\documentclass[12pt,final,showpacs,pra,lengthcheck]{iopart}
\usepackage{ae,aecompl}
\usepackage[T1]{fontenc}
\usepackage[latin9]{inputenc}
\usepackage[active]{srcltx}
\usepackage{xcolor}
\usepackage{textcomp}
\usepackage{amstext}
\usepackage{graphicx}
\PassOptionsToPackage{normalem}{ulem}
\usepackage{ulem}
\usepackage[unicode=true,pdfusetitle,
 bookmarks=true,bookmarksnumbered=false,bookmarksopen=false,
 breaklinks=false,pdfborder={0 0 1},backref=false,colorlinks=true]
 {hyperref}
\usepackage{breakurl}

\makeatletter

\providecolor{lyxadded}{rgb}{0,0,1}
\providecolor{lyxdeleted}{rgb}{1,0,0}

 \@ifundefined{textcolor}{}
 {%
   \definecolor{BLACK}{gray}{0}
   \definecolor{WHITE}{gray}{1}
   \definecolor{RED}{rgb}{1,0,0}
   \definecolor{GREEN}{rgb}{0,1,0}
   \definecolor{BLUE}{rgb}{0,0,1}
   \definecolor{CYAN}{cmyk}{1,0,0,0}
   \definecolor{MAGENTA}{cmyk}{0,1,0,0}
   \definecolor{YELLOW}{cmyk}{0,0,1,0}
 }

\usepackage{textcomp}
\usepackage[english]{babel}

\makeatother

\begin{document}
\title[]{Broadband Excitation by Chirped Pulses: Application to Single Electron Spins in Diamond}

\author{I Niemeyer$^1$, J H Shim$^1$, J Zhang$^1$, D Suter$^1$\footnote{ Author to whom any correspondence should be addressed.}}

\address{$^1$ Fakult\"at Physik, Technische Universit\"at Dortmund, Dortmund, Germany}

\ead{Dieter.Suter@tu-dortmund.de}

\author{T Taniguchi$^2$, T Teraji$^2$, H Abe$^3$, S Onoda$^3$, T Yamamoto$^3$, T Ohshima$^3$, J Isoya$^4$}

\address{$^2$ National Institute for Materials Science, 1-1 Namiki, Tsukuba, Ibaraki, 305-0044 Japan .}

\address{$^3$ Japan Atomic Energy Agency, 1233 Watanuki, Takasaki, Gunma, 370-1292 Japan .}

\address{$^4$ Research Center for Knowledge Communities, University of Tsukuba, Tsukuba, 305-8550 Japan .}

\author{F Jelezko$^5$}

\address{$^5$ Institut f\"ur Quantenoptik, Universit\"at Ulm, Ulm, Germany.}
\begin{abstract}
Pulsed excitation of broad spectra requires very high field strengths
if monochromatic pulses are used. If the corresponding high power
is not available or not desirable, the pulses can be replaced by suitable
low-power pulses that distribute the power over a wider bandwidth.
As a simple case, we use microwave pulses with a linear frequency
chirp. We use these pulses to excite spectra of single NV-centers
in a Ramsey experiment. Compared to the conventional Ramsey experiment,
our approach increases the bandwidth by at least an order of magnitude.
Compared to the conventional ODMR experiment, the chirped Ramsey experiment
does not suffer from power broadening and increases the resolution
by at least an order of magnitude. As an additional benefit, the chirped
Ramsey spectrum contains not only `allowed' single quantum transitions,
but also `forbidden' zero- and double quantum transitions, which can
be distinguished from the single quantum transitions by phase-shifting
the readout pulse with respect to the excitation pulse or by variation
of the external magnetic field strength.
\end{abstract}
\maketitle
\noindent{\it chirpe, broadband, excitation, detection, spin dynamics, quantum computation, quantum information processing, single quantum coherence, double quantum coherence, carbon-13, nearest-neighbor \/}

\pacs{76.30.-v, 76.30.Mi, 76.70.Hb, 33.15.Pw}

\section{Introduction}

Nitrogen-vacancy (NV) defect centers in diamond are promising candidates
for quantum information processing \cite{Jelezko2006}, magnetometry
\cite{JRMaze2008} and electrometry \cite{Dolde2011}. The recently
measured temperature dependence of the zero-field splitting constant
\cite{Acosta2012} indicates that it may also be used as an atomic
temperature sensor. The center consists of a substitutional nitrogen
atom adjacent to a vacancy in the diamond crystal lattice. In the
negatively charged state, it has an electron spin $S=1$. Excitation
with green laser light polarizes the spin at room temperature $\approx90\mbox{ \%}$
\cite{Jelezko2006} into the $\left|m_{s}=0\right\rangle $ ground
state. This state (usually denoted as ``bright state'') exhibits
a higher fluorescence rate than the $\left|m_{s}=\pm1\right\rangle $
spin levels. Microwave pulses can transfer population between the
$\left|m_{s}=0\right\rangle \leftrightarrow\left|m_{s}=\pm1\right\rangle $
spin levels. The populations can be measured via the photon scattering
rate \cite{Jelezko2006}.

Quantum computing with NV-centers can not only use the electron spin,
but also hybrid quantum registers with additional nuclear spins. In
particular, strongly coupled $^{13}\mbox{C}$ nuclear spins have attractive
properties \cite{Dutt2007,Cappellaro2009,Mizuochi2009,Neumann2010,Smeltzer2011}.
The strength of the hyperfine interaction depends on the position
of the nuclear spin \cite{Smeltzer2011} and reaches a maximum of
$130\mbox{ MHz}$ for a $^{13}\mbox{C}$ in a nearest-neighbor lattice
site \cite{Jelezko2006,Felton2009}. Measuring these couplings requires
the recording of spectra that cover a frequency range larger than
the sum of all hyperfine coupling constants. This can be done by ODMR,
which yields spectra with linewidths of several MHz under typical
conditions. These linewidths are the result of power broadening by
the laser and the microwave field. The effect of the laser is eliminated
in the pulsed ODMR approach \cite{Dreau2011}, where the laser is
switched off during the application of the microwave field. The remaining
broadening from the microwave field is also eliminated in the Ramsey
experiments \cite{Ramsey1950,Vion2003}, which yields spectra with
linewidths equal to the natural linewidth. The drawback of the Ramsey
experiment is that it requires excitation pulses that cover the full
bandwidth of the spectrum. This can be challenging for spectra with
large hyperfine couplings.

Here, we present an experimental scheme that avoids power broadening
by using the Ramsey approach of free precession but also avoids the
requirement of strong microwave fields by using excitation pulses
that cover the full bandwidth with very low power. We achieve this
by scanning the frequency over the full spectral range. This type
of pulses are known as chirped pulses \cite{Ferretti1976,Burghardt1990,Jeschke1995}.

Since the microwave field interacts with the different transitions
sequentially, it excites not only the usual, magnetic-dipole allowed
transitions between the $\left|m_{S}=0\right\rangle \leftrightarrow\left|m_{S}=\pm1\right\rangle $
states (single quantum transitions), but also the `forbidden' transition
between the $\left|m_{S}=-1\right\rangle \leftrightarrow\left|m_{S}=+1\right\rangle $
states (double quantum transition). These different types of transitions
can be distinguished by appropriate shifts in the relative phases
of the excitation and readout pulses.

\section{Mathematical Descriptions}

\subsection{Spin S=1/2 System}

\begin{figure}[h]
\begin{centering}
\includegraphics[width=0.5\columnwidth]{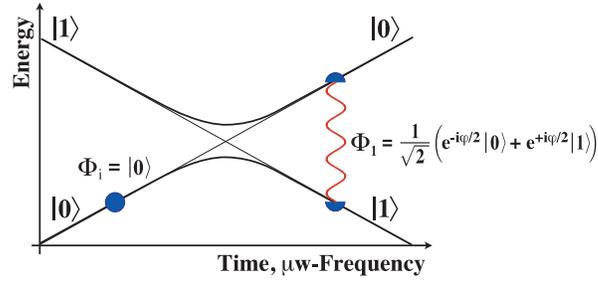}
\par\end{centering}

\caption{Excitation of a two-level system by non-adiabatic rapid passage.\label{fig:Passage}}
\end{figure}

We use chirped excitation pulses to excite transitions in a large
frequency range. Figure\ \ref{fig:Passage} shows the basic idea:
Assuming that we want to excite the transition between the $|m_{S}=0\rangle$
and the $|m_{S}=1\rangle$ state and that the system is initially
in the ground state, we scan the frequency through resonance in such
a way that the system has a $50\,\%$ transition probability to the
$|m_{S}=1\rangle$ state and ends up in the superposition state 
\[
\Phi_{1}=\frac{1}{\sqrt{2}}\left(e^{-i\varphi_{1}/2}|0\rangle+e^{i\varphi_{1}/2}|1\rangle\right),
\]
which maximizes the coherence between the two levels. The relative
phase $\varphi$ depends on the phase, amplitude and scan rate of
the microwave. 

The effect of the chirped pulse can thus be described by a unitary
operator \cite{Jeschke1995}
\[
U_{1}=e^{-i\varphi_{1}S_{z}}e^{-i\frac{\pi}{2}S_{y}}.
\]

\begin{figure}[h]
\begin{centering}
\includegraphics[width=0.55\columnwidth]{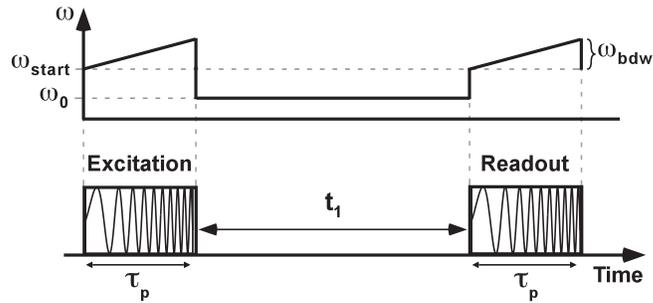}
\par\end{centering}

\caption{Pulse sequence for broadband Ramsey experiment with chirped excitation
pulses. $\omega_{start}$ defines the start frequency of the scan
and $\omega_{bdw}$ the width of the scan. $\omega_{0}$ is the reference
frequency that relates the phase of the two pulses; for details see
text. $\tau_{\mbox{\tiny p}}$ is the pulse duration and $t_{1}$
the free evolution time which is incremented between experiments.
\label{fig:PulsSequence}}
\end{figure}

As shown in figure\ \ref{fig:PulsSequence}, the system is then allowed
to evolve freely for a time $t_{1}$. If $\Omega_{0}$ is the Larmor
frequency of the system, the superposition state acquires an additional
phase $\Omega_{0}t_{1}$ during this time. The resulting state is
\begin{eqnarray*}
\Phi_{2} & = & e^{-i\Omega_{0}t_{1}S_{z}}\Psi_{1}\\
 & = & \frac{1}{\sqrt{2}}\left(e^{-i(\Omega_{0}t_{1}+\varphi_{1})/2}|0\rangle+e^{i(\Omega_{0}t_{1}+\varphi_{1})/2}|1\rangle\right).
\end{eqnarray*}
 At this point, a second chirped pulse generates another transformation
that we write as 
\[
U_{2}=e^{-i\frac{\pi}{2}S_{y}}e^{-i\varphi_{2}S_{z}},
\]
thus converting the system into the final state 
\begin{eqnarray*}
\Phi_{3} & = & i\sin\left(\frac{\Omega_{0}t_{1}+\varphi_{1}+\varphi_{2}}{2}\right)\left|0\right\rangle \\
 &  & +\cos\left(\frac{\Omega_{0}t_{1}+\varphi_{1}+\varphi_{2}}{2}\right)\left|1\right\rangle 
\end{eqnarray*}
The population of the ground/bright state $|0\rangle$ is thus

\begin{eqnarray*}
P(\left|0\right\rangle ) & = & \left[\sin\left(\frac{\Omega_{0}t_{1}+\varphi_{1}+\varphi_{2}}{2}\right)\right]^{2}\\
 & = & \frac{1}{2}\left[1-\cos\left(\Omega_{0}t_{1}+\varphi_{1}+\varphi_{2}\right)\right].
\end{eqnarray*}
Clearly, this corresponds to a Ramsey-fringe pattern, which can be
Fourier-transformed to obtain the spectrum (a single line at $\Omega_{0}$
in this case).

\subsection{Spin S=1 System}

\begin{figure}[h]
\begin{centering}
\includegraphics[width=0.55\columnwidth]{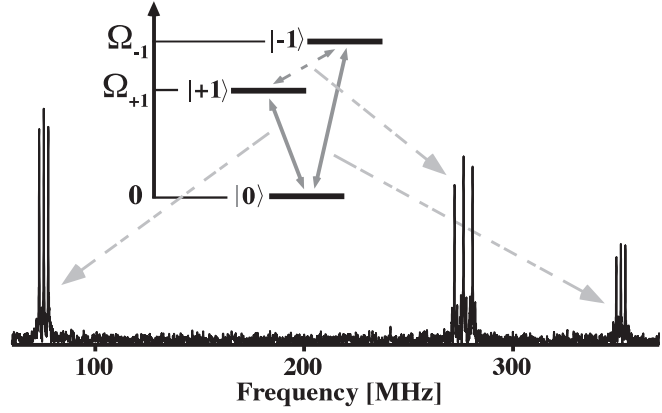}
\par\end{centering}

\caption{Relevant three-level system. The full arrows indicate allowed magnetic
dipole transitions.\label{fig:Spin1}}
\end{figure}

The NV-center in diamond is a spin $S=1$ system. We write the relevant
Hamiltonian
\begin{equation}
\mathcal{H}=DS_{z}^{2}+\Omega_{0}S_{z}.\label{eq:HamEl}
\end{equation}
Here, $D=2.8$ GHz is the zero-field splitting and $\Omega_{0}$ the
Larmor frequency due to the interaction with the magnetic field. Figure\ \ref{fig:Spin1}
shows the resulting level structure, together with the allowed magnetic
dipole transitions, marked by arrows. We write $|m_{S}\rangle$ for
the eigenstates of the Hamiltonian, where $m_{S}$ is the eigenvalue
of $S_{z}$.

In the following, we assume that the Rabi frequency is small compared
with the frequency separation of the relevant transitions. We therefore
can assume that the microwave field drives only one transition at
a time \cite{Ferretti1976,Burghardt1990,Jeschke1995}. If we scan
from low to high frequency, we first excite the transition $|0\rangle\leftrightarrow|+1\rangle$
in the system shown in figure\ \ref{fig:Spin1}. Starting from the
initial state $\Psi_{0}=|0\rangle$, the first passage through resonance
converts it into

\begin{eqnarray*}
\Psi_{1} & = & U_{zy}\left(\varphi,\theta\right)\left|0\right\rangle \\
 & = & e^{i\varphi/2}\cos\frac{\theta}{2}\left|0\right\rangle -e^{-i\varphi/2}\sin\frac{\theta}{2}\left|+1\right\rangle ,
\end{eqnarray*}
where $\theta$ is the effective flip-angle of the pulse. Passing
through the second resonance, we obtain
\begin{eqnarray*}
\Psi_{2} & = & U_{zy}\left(\varphi,\theta\right)\Psi_{1}\\
 & = & -\sin\frac{\theta}{2}\cos\frac{\theta}{2}\left|-1\right\rangle -e^{-\mbox{i}\varphi/2}\sin\frac{\theta}{2}\left|+1\right\rangle \\
 &  & +e^{\mbox{i}\varphi}\cos^{2}\frac{\theta}{2}\left|0\right\rangle .
\end{eqnarray*}
Here, we have assumed that the effect of the pulse on both transitions
is the same. This is a good approximation if the scan rate and the
transition strengths are the same.

During the subsequent free evolution period, the system evolves to

\begin{eqnarray*}
\Psi_{3} & = & U_{z}\left(t_{1}\right)\Psi_{2}\\
 & = & -e^{-i\left(\Omega_{-1}t_{1}\right)}\sin\frac{\theta}{2}\cos\frac{\theta}{2}\left|-1\right\rangle \\
 &  & -e^{-i\left(\Omega_{+1}t_{1}+\varphi/2\right)}\sin\frac{\theta}{2}\left|+1\right\rangle +e^{i\varphi}\cos^{2}\frac{\theta}{2}\left|0\right\rangle ,
\end{eqnarray*}
with $\Omega_{\pm1}=D\mp\Omega_{0}$ representing the resonance frequencies
of the two transitions. 

This free precession period is terminated by the readout pulse, which
is identical to the excitation pulse (apart from an overall phase).
It converts part of the coherences back to populations. Here, we are
interested only in the population $P_{0}=P(\left|0\right\rangle )$
of the bright state $\left|0\right\rangle $:

\begin{eqnarray*}
P_{0} & = & \left|A_{1}\left(e^{-\mbox{i}\left(\Omega_{-1}t_{1}+\frac{\varphi}{2}\right)}+e^{-\mbox{i}\left(\Omega_{+1}t_{1}+\frac{\varphi}{2}\right)}\right)+A_{2}e^{\mbox{i}2\varphi}\right|^{2}\\
 & = & 2A_{1}^{2}+A_{2}^{2}\\
 & + & 2A_{1}^{2}\cos\left(\left[\Omega_{+1}-\Omega_{-1}\right]t_{1}\right)\\
 & + & 2A_{1}A_{2}\left[\cos\left(\Omega_{+1}t_{1}+\frac{5\varphi}{2}\right)+\cos\left(\Omega_{-1}t_{1}+\frac{5\varphi}{2}\right)\right],
\end{eqnarray*}
with the amplitudes
\[
A_{1}=\sin^{2}\frac{\theta}{2}\cos\frac{\theta}{2}\quad,\quad A_{2}=\cos^{4}\frac{\theta}{2}.
\]
The first term in this expression is a constant offset. The second
term oscillates at the frequency $2\Omega_{0}=\Omega_{-1}-\Omega_{+1}$
of the $|-1\rangle\leftrightarrow|+1\rangle$ transition, while the
third term contains the two single quantum transition frequencies.
Fourier transformation of this will therefore yield a spectrum with
the two allowed single quantum transition and the `forbidden' double
quantum transition frequency, as shown in figure\ \ref{fig:Spin1}.
Note that the frequencies in the figure are not the true resonance
frequencies. The relation between the apparent and the real frequencies
will be discussed in the following section.

\section{Experimental Results}

\subsection{Setup and Samples}

The experiments were performed with a home-built confocal microscope.
A diode-pumped solid-state laser with an emission wavelength of $532\mbox{ nm}$
was used. The cw laser beam was sent through an acousto-optical modulator
to generate laser pulses for excitation and readout. We used an oil
immersion microscope objective (with $\mbox{NA}=1.4$) mounted on
a nano-positioning system to focus the laser light to single NV-centers.
The microscope objective also collects light emitted by the NV-centers
during readout. For electronic excitation we used a setup consisting
of a microwave synthesizer and an arbitrary waveform generator, which
were connected to a mixer and up-converted. Here the synthesizer was
used as local oscillator and the arbitrary waveform generator, which
had a sampling frequency of $4\mbox{ GS/s}$, delivered the intermediate
frequency. We were able to control the phase as well as the frequency
of the up-converted signal by changing the phase and the frequency
of the arbitrary waveform generator. The controllable frequency bandwidth
was $<2\mbox{ GHz}$. The microwaves were guided through a Cu wire
mounted on the surface of the diamond. The maximal excitation power
was $8\mbox{ W}$. We used a permanent magnet to apply a magnetic
field to the sample.

We applied the chirped Ramsey sequence shown in figure\ \ref{fig:PulsSequence}
to two different diamond samples both of type IIa. One is a $^{12}\mbox{C}$
enriched (concentration of $99.995\,\%$) diamond with a relaxation
time of $T_{2}^{*}>200\mbox{ \ensuremath{\mu}s}$ the other a natural
abundance diamond with $T_{2}^{*}\approx1\mbox{ \ensuremath{\mu}s}$.

The enriched sample is a diamond single crystal grown at $5.5\mbox{ GPa}$
and $1400\mbox{ \textdegree C}$ from Co-Ti-Cu alloy by using a temperature
gradient method. As a solid carbon source, polycrystalline diamond
plates synthesized by chemical vapor deposition (CVD) utilizing $^{12}\mbox{C}$
enriched methane were used. Secondary ion mass spectrometry (SIMS)
analysis has shown that typically a $^{12}\mbox{C}$ concentration
of $99.995\text{ \%}$ in the grown crystals was achieved. The crystal
was irradiated at room temperature with 2 MeV electrons and a total
flux intensity of $10^{11}/\mbox{cm}^{2}$. Subsequently it was annealed
at $1000\text{ \textdegree C}$ for 2 hours in vacuum.

We first present measurements of the enriched sample to illustrate
different features of this experiment, in particular how the phases
of the excitation pulses affect the observed frequency and phase of
the different types of resonance lines.

\subsection{Reference Frequency}

In the experiments, we are not interested in the dc component $2A_{1}^{2}+A_{2}^{2}$,
which we omit in the following. We now compare experiments where we
change the phase of the second pulse with respect to that of the first
one by an angle $\alpha$. The resulting signal is then

\begin{eqnarray}
s & = & 2A_{1}^{2}\cdot\cos\left(\left[\Omega_{+1}-\Omega_{-1}\right]t_{1}\right)\nonumber \\
 &  & +2A_{1}A_{2}\left[\sin\left(\Omega_{-1}t_{1}+\frac{5}{2}\varphi-\alpha\right)\right.\nonumber \\
 &  & \left.+\sin\left(\Omega_{+1}t_{1}+\frac{5}{2}\varphi-\alpha\right)\right]\label{eq: Signal_PhaseSensitveReadout_AnalyticalFormular}
\end{eqnarray}
In the experiments, we use this additional phase for two purposes:
we increment it linearly with the free precession period $t_{1}$
to shift the effective precession frequency, and we use it to distinguish
the double quantum transition, which does not depend on $\alpha$,
from the single quantum transitions.

Looking first at the linear phase increments, we set $\alpha=\omega_{0}t_{1}$.
The resulting signal is then 

\begin{eqnarray*}
s_{1} & = & 2A_{1}^{2}\cdot\cos\left(\left[\Omega_{+1}-\Omega_{-1}\right]t_{1}\right)\\
 &  & +2A_{1}A_{2}\left[\sin\left(\left(\Omega_{-1}-\omega_{0}\right)t_{1}+\frac{5}{2}\varphi\right)\right.\\
 &  & \left.+\sin\left(\left(\Omega_{+1}-\omega_{0}\right)t_{1}+\frac{5}{2}\varphi\right)\right].
\end{eqnarray*}

\begin{figure}[h]
\begin{centering}
\includegraphics[width=0.65\columnwidth]{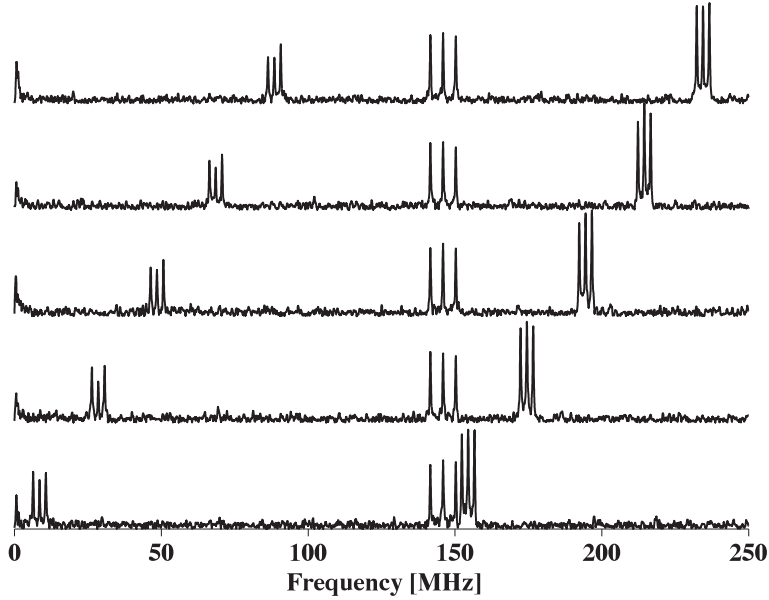}
\par\end{centering}

\caption{\label{fig: Absolute Spectra for different detunings} Ramsey spectra
measured with different reference frequencies. The actual transition
frequencies are: $\Omega_{+1}/2\pi=2798.5\mbox{ MHz}$, $\Omega_{-1}/2\pi=2944.5\mbox{ MHz}$
and $2\Omega_{0}/2\pi=146\mbox{ MHz}$. The reference frequencies
were: $\omega_{0}/2\pi=2790\mbox{, }2770\mbox{, }2750\mbox{, }2730\mbox{, }2710\mbox{ MHz}$
(from bottom to top). For all spectra, the start frequency of the
chirp was $2770\mbox{ MHz}$ and the width 250 MHz.}

\end{figure}

We therefore expect that the single quantum transitions appear shifted
to the frequencies $(\Omega_{\pm1}-\omega_{0})$, while the double
quantum transition remains at the natural frequency $2\Omega_{0}=\Omega_{+1}-\Omega_{-1}$.
This is clearly borne out in figure\ \ref{fig: Absolute Spectra for different detunings},
where we compare spectra obtained with the same excitation scheme,
but different reference frequencies. The three groups of lines appear
centered around $\Omega_{+1}-\omega_{0}$, $2\Omega_{0}=\Omega_{+1}-\Omega_{-1}$,
and $\Omega_{-1}-\omega_{0}$. For these experiments, we chose $\omega_{0}$
such that the resulting frequencies fall into a frequency window that
is easily accessible. In the case of the spectra shown here, we incremented
$t_{1}$ by $2\mbox{ ns}$ between scans, which yields, according
to the Nyquist theorem a $250\mbox{ MHz}$ frequency window. The maximum
value of $t_{1}$ was $5\mbox{ \ensuremath{\mu}s}$. The data were
recorded in the same magnetic field, which splits the $\left|m_{s}=\pm1\right\rangle $
lines by $146\mbox{ MHz}$. All measurements were done with frequency
chirps starting at $2770\mbox{ MHz}$ and the pulse lengths were $\tau_{p}$
= $120\mbox{ ns}$. It is clearly seen that the single quantum transitions
are shifted in the opposite direction from the reference frequency,
while the double quantum transitions (at $146\mbox{ MHz}$) are not
affected by the detuning.

\subsection{Phase Shifts}

\begin{figure}[h]
\begin{centering}
\includegraphics[width=0.6\columnwidth]{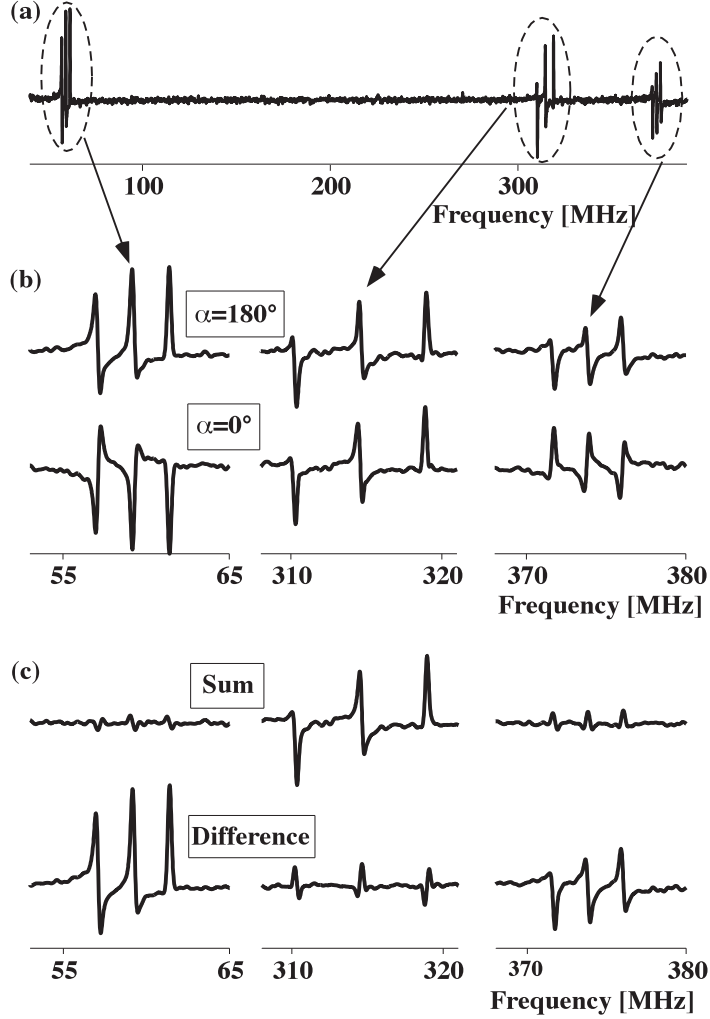}
\par\end{centering}

\caption{\label{fig:Phase-spectra-of-two-chirped-Ramseys-1} Phase-sensitive
spectra of two chirped Ramsey measurements. (a) Full spectrum. (b)
Real parts of spectra obtained with phase shifts $\alpha=180{^\circ}$
(top) and $\alpha=0{^\circ}$ (bottom). (c) Sum (top) and difference
(bottom) of the spectra in (b). }
\end{figure}

Instead of incrementing the phase proportionally with $t_{1}$, we
can also compare two spectra with different constant phase shifts
of the readout pulse. The two traces of figure\ \ref{fig:Phase-spectra-of-two-chirped-Ramseys-1}\ (b)
show an example: the spectra were obtained with phase shifts of $0$
and $\pi$ between the two pulses; only expanded regions of the full
spectrum shown in figure\ \ref{fig:Phase-spectra-of-two-chirped-Ramseys-1}\ (a)
are shown. These data were recorded with a different NV-center in
a higher magnetic field strength. The chirp bandwidth was $500\mbox{ MHz}$,
the pulse length $\tau_{p}$ = $50\mbox{ ns}$ and the maximum value
of $t_{1}$ was $5\mbox{ \ensuremath{\mu}s}$. According to equation\ (\ref{eq: Signal_PhaseSensitveReadout_AnalyticalFormular}),
we expect that the phase of the single quantum transitions $|0\rangle\leftrightarrow|\pm1\rangle$
should change with $\alpha$, while the double quantum transition
$|+1\rangle\leftrightarrow|-1\rangle$ should not change. Inspection
of the experimental data shows that the spectral lines close to $60$
and $375\mbox{ MHz}$ are inverted between the two spectra, while
the signals close to $315\mbox{ MHz}$ do not change. We therefore
interpret the outer lines as single quantum transitions, the inner
ones as double quantum transitions. This assignment is also consistent
with the splittings due to the hyperfine interaction with the $^{14}$N
nuclear spin, which is $2.15\mbox{ MHz}$ for the single quantum transitions
and $4.3\mbox{ MHz}$ for the double quantum transition.

Using this phase dependence, we can also separate the two types of
transitions by calculating the sum and difference of the two spectra.
According to equation\ (\ref{eq: Signal_PhaseSensitveReadout_AnalyticalFormular}),
the difference of the two spectra should be 

\begin{eqnarray}
s_{\alpha=0\text{\textdegree}}-s_{\alpha=180\text{\textdegree}} & = & 4A_{1}A_{2}\left[\sin\left(\Omega_{-1}t_{1}+\frac{5}{2}\varphi\right)\right.\nonumber \\
 &  & \left.+\sin\left(\Omega_{+1}t_{1}+\frac{5}{2}\varphi\right)\right],\label{eq: Phase-Sens-Readout_DIFF}
\end{eqnarray}
and the sum

\begin{eqnarray}
s_{\alpha=0\text{\textdegree}}+s_{\alpha=180\text{\textdegree}} & = & 4A_{1}^{2}\cdot\cos\left(2\Omega_{0}t_{1}\right).\label{eq: Phase-Sens-Readout_SUM}
\end{eqnarray}

The lower part of figure\ \ref{fig:Phase-spectra-of-two-chirped-Ramseys-1}
shows the result of this operation: The sum (upper trace) contains
mostly the double quantum signals, while the difference is dominated
by the single quantum transitions which corresponds to the results
of equation\ (\ref{eq: Phase-Sens-Readout_DIFF}) and (\ref{eq: Phase-Sens-Readout_SUM}).
The incomplete suppression of the other signals can be attributed
to instabilities in the experimental setup, which result in thermal
frequency shifts and changing amplitudes.

\subsection{B-Field Dependence}

\begin{figure}[h]
\begin{minipage}[t]{1\columnwidth}%
\begin{center}
\includegraphics[width=0.6\columnwidth]{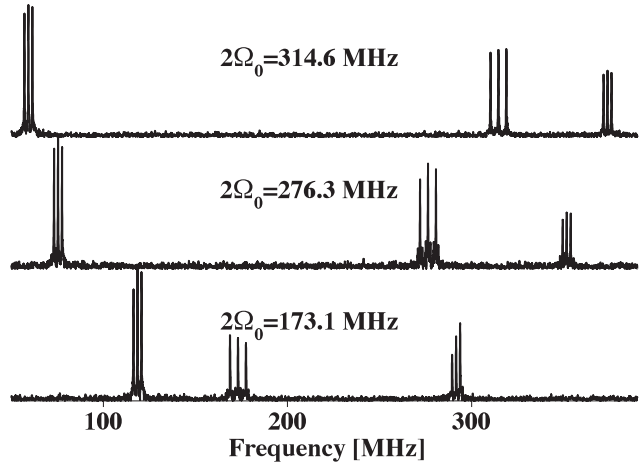}
\par\end{center}

\caption{\label{fig: Chirped-Ramsey-Spectra-BFiedScan} Absolute value spectra
for different magnetic field strengths. $2\Omega_{0}$ corresponds
to the separation of the $\left|+1\right\rangle ,\,\left|-1\right\rangle $
levels and therefore to the separation between the two single quantum
transitions and to the center frequency of the double quantum transition
(inner line of the triplet). }
\end{minipage}
\end{figure}

Figure\ \ref{fig: Chirped-Ramsey-Spectra-BFiedScan} shows spectra
of the $^{12}$C enriched crystal for different magnetic field strengths.
For these measurements the reference frequency was $\omega_{0}=2670.8\mbox{ MHz}$.
The chirp pulses had a bandwidth of $500\mbox{ MHz}$ and a duration
of $\tau_{p}$ = $50\mbox{ ns}$. The start frequency of the chirp
was $\omega_{start}=2650.8\mbox{ MHz}$ and the bandwidth $\omega_{bdw}=500\mbox{ MHz}$.
The sampling interval of 1 ns results in a bandwidth of $500\mbox{ MHz}$
and maximum value of $t_{1}$ of $5\mbox{ \ensuremath{\mu}s}$ yields
a digital frequency resolution of $100\mbox{ kHz}$. 

In each spectrum of the figure, we list the splitting between the
single quantum transitions, which corresponds to the magnetic field
component along the symmetry axis of the center, measured in frequency
units. The outer triplets correspond to the single quantum transitions
($\left|0\right\rangle \leftrightarrow\left|\pm1\right\rangle $),
the inner lines to the double quantum transition ($\left|+1\right\rangle \leftrightarrow\left|-1\right\rangle $).
With increasing magnetic field strength, the splitting between the
single quantum transitions increases proportionally and is always
equal to the frequency of the double quantum transition. The frequency
changes for the left and right triplets are not the same, this can
be explained by transversal components in the Zeeman interaction which
we have neglected in the Hamiltonian equation\ (\ref{eq:HamEl}).

\subsection{Multi-Line Broadband Spectrum}

The chirped excitation scheme is particularly useful when the spectra
cover a broad frequency range with many resonance lines. Such a situation
exists in NV-centers with a $^{13}$C nuclear spin in the first coordination
shell.

\begin{figure}[h]
\begin{minipage}[t]{1\columnwidth}%
\begin{center}
\includegraphics[width=0.65\columnwidth]{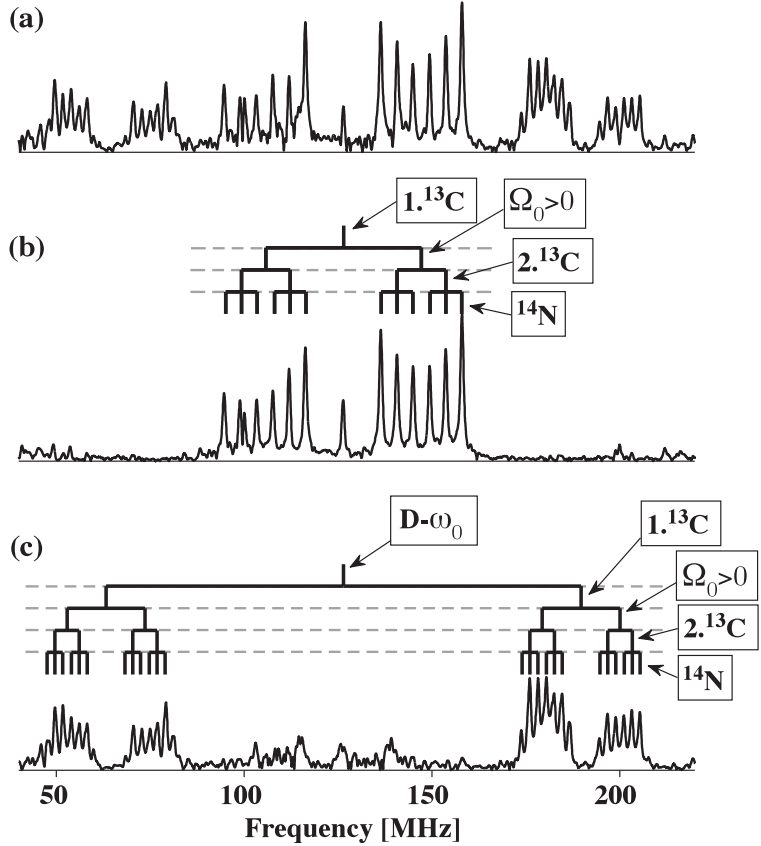}
\par\end{center}

\caption{\label{fig: Chirped-Ramsey_NaturalAbondance_13CNextNeighbor_LowNoise}
Spectra of NV-center in natural abundance diamond with two adjacent
$^{13}\mbox{C}$ nuclear spins. One strongly coupled with $A_{\Vert}\approx126.5\mbox{ MHz}$
(nearest-neighbor) and one with $A_{\Vert}\approx6.55\mbox{ MHz}$
\cite{Smeltzer2011}. $\Omega_{0}\approx10\mbox{ MHz}$ is the Zeeman
interaction, $D$ the zero-field splitting and $\omega_{0}$ the reference
frequency. (a) Absolute value spectrum. (b) sum and (c) difference
of the spectra obtained with phase shifts $\alpha=0{^\circ}$ and
$\alpha=180{^\circ}$. }
\end{minipage}
\end{figure}

figure\ \ref{fig: Chirped-Ramsey_NaturalAbondance_13CNextNeighbor_LowNoise}
shows the spectrum of such a center. In this particular center, the
electron spin is coupled to a nearest-neighbor $^{13}$C nuclear spin
with a hyperfine coupling constant $A_{\|}\approx126.5\mbox{ MHz}$
as well as to an additional $^{13}\mbox{C}$ with a coupling constant
of $A_{\|}\approx6.55\mbox{ MHz}$. For this measurement we used a
type IIa natural abundance diamond and applied a magnetic field strength
of approximately $9\mbox{ G}$. The field was not aligned and had
an angle of $\approx65\text{\textdegree}$ with respect to the symmetry
axis of the NV-center, which corresponded to a projected field strength
of $3.7\mbox{ G}$. The chirp bandwidth was $250\mbox{ MHz}$, starting
from $2750.3\mbox{ MHz}$ and the pulse-duration was $\tau_{p}$ =
$60\mbox{ ns}$.

The top graph of figure\ \ref{fig: Chirped-Ramsey_NaturalAbondance_13CNextNeighbor_LowNoise}
shows the absolute value of a chirped Ramsey spectrum. The center
graph shows the sum and the lower the difference of two phase-shifted
spectra, which correspond to the double- and single quantum transitions,
respectively. The line at 126.5 MHz in b) is a zero-quantum transition.
Its transition frequency matches the hyperfine coupling constant of
the nearest-neighbor $^{13}$C. In the spectra, we also indicate how
the spectral lines can be assigned to transitions of the electron
spin with different configurations of the three coupled nuclear spins.
If we consider only the Hamiltonian of equation\ (\ref{eq:HamEl})
for the electron spin and the hyperfine interactions with the nuclear
spins, the single quantum spectrum (bottom of figure\ \ref{fig: Chirped-Ramsey_NaturalAbondance_13CNextNeighbor_LowNoise})
should consist of 4 groups of six lines. In the experimental spectrum,
the four groups contain more than six lines. This difference can be
attributed to the splitting of the $\left|m_{S}=0\right\rangle $
ground state due to the interaction with the transverse components
of the magnetic field and the nonsecular hyperfine interaction.\cite{DetailsElsewhere}

\section{Conclusions}

We have introduced a new experimental technique for measuring broad
spectra of single electron spins. This approach does not require high
microwave power. The precession frequency of the spins is measured
in the absence of microwave irradiation, in the form of Ramsey fringes,
which results in high resolution spectra. The resulting spectra contain
not only the dipole-allowed single quantum transitions, but also multiple
quantum transitions that can only be excited by multiple absorption/emission
processes. This technique is particularly useful in the case of electron
spins coupled to multiple nuclear spins. Such clusters of spins may
be useful tools for quantum computing applications \cite{Dutt2007,Cappellaro2009,Mizuochi2009,Neumann2010}.
We have demonstrated the technique on the example of single electron
spins in the diamond NV-center, but the same approach should also
be applicable to other systems, where the excitation bandwidth can
be sufficiently large.

\ack

This work was supported by the Deutsche Forschungsgesellschaft through
grant Su 192/27-1 (FOR 1482).

\section*{References}

\bibliographystyle{unsrt}
\bibliography{ChirpedRamsey}

\end{document}